\documentstyle{article}
\begin{document}
\LARGE
\begin{center}
\bf A Proposal of Testing Imaginary Time in a Total Reflection
\vspace*{0.6in} \normalsize \large \rm

Zhong Chao Wu

Center for astrophysics

University of Science and Technology of China

Hefei, Anhui, China

(Oct. 16, 1997)

\vspace*{0.25in}

published in  \it Journal of University of Science and Technology of
China, \rm No. 5,  4-7 (1998)

\vspace*{0.8in} \large \bf Abstract
\end{center}
\vspace*{.5in} \rm \normalsize

All paradoxes concerning faster-than-light signal propagation
reported in recent experiments can be dispelled by using
imaginary time in a quantum framework. I present a proposal of
testing imaginary time in a total reflection.

\vspace*{0.95in}


\vspace*{0.25in} \pagebreak

There have been many claims that a particle may travel faster than
light [1]. It seems to contradict the principle of relativity. One
of the motivations of this letter is to clarify this point. I shall
work in a quantum framework. Quantum theory has been refined
significantly by new developments in quantum cosmology [2][3]. A
very brief review is helpful.

In quantum cosmology, the wave function of a closed universe
usually take some superposition form of the WKB wave packet:
\begin{equation}
\Psi \approx C\exp (-S/\hbar),
\end{equation}
where $C$ is a slowly varying prefactor and $S \equiv S_r + i
S_i$ is the complex phase and time coordinate does not appear
explicitly.

Inserting this into the zero energy Schroedinger  or  the
Wheeler-DeWitt equation, one obtains
\begin{equation}
\left(- \frac{1}{2} (\bigtriangledown S)^2 + V  + \hbar \left(
\frac{1}{2}
\bigtriangleup S + \bigtriangledown S \cdot \bigtriangledown
\right) - \frac{1}{2}\hbar^2 \bigtriangleup\right ) C  = 0,
\end{equation}
where the operators $\bigtriangleup$ and $\bigtriangledown$ are
with respect to the supermetric of the configuration space, $V$
is the potential. We
can separate this equation into a real part
\begin{equation}
\left(-\frac{1}{2}(\bigtriangledown S_r)^2 +
\frac{1}{2}(\bigtriangledown S_i)^2 + V + \hbar \left(\frac{1}{2}
\bigtriangleup S_r + \bigtriangledown S_r \cdot \bigtriangledown
\right)  - \frac{1}{2}\hbar^2\bigtriangleup \right ) C= 0,
\end{equation}
and an imaginary part
\begin{equation}
-C \bigtriangledown S_r \cdot \bigtriangledown S_i +
\hbar\left(\frac{1}{2}C \bigtriangleup S_i + \bigtriangledown S_i
\cdot \bigtriangledown C\right) = 0
\end{equation}

If we ignore the quantum effects represented by the terms
associated with powers of $\hbar$ in these equations, then
Eqs (3) and (4) become
\begin{equation}
-\frac{1}{2} (\bigtriangledown S_r)^2 + \frac{1}{2} (
\bigtriangledown S_i)^2 + V = 0,
\end{equation}
and
\begin{equation}
\bigtriangledown S_r \cdot \bigtriangledown S_i = 0.
\end{equation}

If $\bigtriangledown S_r = 0$, then Eq. (5) is the
Lorentzian Hamilton-Jacobi equation, with $S$ and
$\bigtriangledown S_i$ identified as the classical action and
the canonical momenta respectively. One can define Lorentzian
orbits along integral curves with $\frac{\partial}{\partial t}
\equiv \bigtriangledown S_i \cdot \bigtriangledown$. The wave
function represents an ensemble of classical trajectories.

Even if $\bigtriangledown S_r \neq 0$, one can still consider
Eq.(5) as the modified Hamilton-Jacobi equation. However, the
Lorentzian trajectories will trace out orbits in the presence of
the potential $V - \frac{1}{2} (\bigtriangledown S_r)^2$. If the
modification is not negligible, then the Lorentzian evolutions
should deviate quite dramatically from classical dynamics.
On the other hand, from Eq.(6) we know that $S_r$
remains constant along these orbits and that $\exp (-S_r)$
can be interpreted as the relative probability of these
trajectories.

One may also interpret this situation as follows: The wave
function consists of two parts of dynamics. $S_i$ represents the
Lorentzian evolutions in real time, while $S_r$ represents
the Euclidean evolutions in imaginary time. Sometimes one can
rephrase the situation by saying that the second part of dynamics
is frozen from the viewpoint of real time. The evolutions in real
time is causal, while the evolutions in imaginary time is
stochastic. This becomes a common phenomenon in quantum
mechanics. Indeed, the results of recent experiments in quantum
optics concerning tunneling time can be thought of as the first
experiment confirmation of the existence of imaginary time [4].
The second motivation of this letter is to present a experimental
test of imaginary time in a total reflection.

If $\bigtriangledown S_r = 0$, then eq. (4) implies the
probability conservation
of the Lorentzian evolutions. The $\hbar$ term in (3)
represents the creation of probability during the Euclidean
evolution, it is intertwined with the dynamics in imaginary time.
The nonconservation of probability in imaginary time is
consistent with the scenario of the birth of the universe.

The above argument can be applied even to ordinary systems
governed by the Schroedinger equation. For a system with a
time-independent Hamiltonian, solutions can be decomposed in
terms of the stationary states satisfying the truncated equation,
where the external time disappears. One can view the truncated
equation as the Wheeler-DeWitt equation. The intrinsic time of
the system will emerge naturally from the wave packets as
described above [4].

It is believed that in general information is propagated with
the group velocity [5]. The velocities  measured in the relevant
experiments mentioned in this paper are associated with the
motion in classically forbidden region, and group velocity does
not make sense then.

One can easily make a model in this quantum framework. Assume a
particle with mass $m$ is moving in a box, i.e. which
potential energy $U = 0$ for $0<x<a, 0< y <b, -\infty < z <
\infty$ and $U = \infty$ elsewhere. The truncated Schroedinger
equation for a stationary state with energy $E$ is
\begin{equation}
\bigtriangleup \psi + (2m/\hbar^2) E\psi = 0,
\end{equation}
in the box. The wave function $\psi$ takes the form of the
product:
\begin{equation}
\psi_{n_1 n_2 k} \sim \sin \frac{\pi n_1}{a}x\sin \frac{\pi
n_2}{b}y\exp {ikz},
\end{equation}
where $n_1, n_2$ are integers and $k$ is a complex number
satisfying
\begin{equation}
E = \frac{\pi^2 \hbar^2}{2m} \left (
\frac{n_1^2}{a^2}+\frac{n_2^2}{b^2} + \frac{k^2}{\pi^2}\right ).
\end{equation}
Depending on whether $k$ is real or imaginary, the wave
function associated with the motion along $z$-direction would
take oscillatory or exponential form. The sign of $k$ chosen
depends on the way to feed particles into the box from its ends
$z = \pm \infty$.

If $k$ is imaginary, then according to our philosophy the
time associated with this degree of freedom becomes imaginary,
while the time associated with the motion along $x$ and $y$
directions always remains real. Under this circumstance, it takes
no real time for a particle to travel along $z$ direction. The
imaginary time lapse causes the wave function to decay, or to
reduce the probability density.

It is well known that in a guided wave, for a given mode $i$ of
vibration, there is a lowest possible frequency, the critical
frequency $\omega_i$. All modes with frequencies below this
threshold should be damped out. Then one can state that wave
propagation undergoes nonzero imaginary and zero-real time
intervals. Indeed, the guided wave can be considered as a
coherent oscillation of an ensemble of photons. The Schroedinger
equation (7) with $2mE/\hbar^2$ replaced by $\omega^2$ behaves as
a sort of Maxwell wave equation
for the guided wave. To measure the time lapse from outside the
waveguide, people may get an illusion that the wave propagates at
a speed higher than that of light. However, we are only concerned
about the local speed of light. Here the principle of
relativity concerning the speed of light in vacuum is not
violated. The
origin of the illusion that the signal travels at a speed higher
than that of light is that the classically forbidden region is
shortcutted by the imaginary time. This has been confirmed
experimentally [6] [7].

It is interesting to write out the phase velocity $v_p$
and group velocity $v_g$ of the guided wave with the frequency
higher than the threshold (we set $c = 1$) [5]:
\begin{equation}
v_p = \frac{\omega}{k} = \frac{\sqrt{k^2 + \omega_i^2}}{k},\;\;
v_g = \frac{d\omega}{dk}=\frac{k}{\sqrt{k^2 + \omega_i^2}}.
\end{equation}
In real time the group velocity or the velocity of signal
propagation approaches zero when the frequency is close to the
threshold. However, when the frequency is lower than the
threshold, then the group velocity does not make sense and the
signal velocity tends to infinity.

In some reports, we may learn that some signal travels
across space at a speed of several times the speed of light. The
numerical value is not essential since it is obtained through the
average of the infinite value in the imaginary time zone and 1 or
less in the real time zone.

I would like to propose a new proposal of a experimental test.
In an internal reflection a light strikes the boundary $ z = 0$
from the side $ z < 0$ of a optically denser medium with
index $n_1$ to a less dense medium with index $n_2$. We assume
the light travels along the $xz$ plane. The wave vector $\bf
k$ \rm satisfies
\begin{equation}
k_{0x} = k_{1x} = k_{2x},\;
k_{1z}= - k_{oz} = - \frac{\omega} n_1 \cos \theta_0,\;
k_{2z} = \omega \sqrt{n^2_2 - n_1^2 \sin^2 \theta_0}.
\end{equation}
where $\omega$ is the frequency of the light, $\theta_o$ is the
angle of incidence. The angle of refraction $\theta_2$ is
obtained
\begin{equation}
\frac{\sin \theta_2}{\sin \theta_0} = \frac{n_1}{n_2}.
\end{equation}
A total reflection occurs if $\sin \theta_2 \geq 1$, then
$\theta_2$ and $k_{2z}$ become imaginary. The light in the less
dense medium will travel along the $x$ direction and decay
exponentially along the positive $z$ direction. In a similar way
as in the guided wave case, the decay light along the $z$
direction can be considered as a propagation in imaginary time,
the observer will find that it takes zero real time for the light
to cross medium 2.

In summary, one can interpret all claims on the signal
propagation with a speed higher than the speed of light as a
particle traveling in a classically forbidden region in imaginary
time. Relativity is intact.

\rm
\normalsize

\begin{center}
\vspace*{.05in} \large \bf References
\end{center}
\vspace*{.05in} \rm \normalsize

[1]. R.Y. Chiao, \it Phys. Rev. \rm \bf A 48, \rm R34 (1993).
R.Y. Chaio, P.G. Kwait and A.M. Steinberg, \it Scientific
American, \rm 52, August 1993. R.Y. Chiao and J. Boyce, \it Phys.
Rev. Lett. \rm \bf 73, \rm 3383 (1994).

[2]. G.W. Gibbons and S.W. Hawking, \it Euclidean Quantum
Gravity, \rm (World Scientific, Singapore, 1992).

[3]. S.W. Hawking, \it Nucl. Phys. \bf B239, \rm 257 (1984).

[4]. Zhong Chao Wu, \it The Imaginary Time in the Tunneling Process.
\rm (unpublished, 1995).

[5]. L. Brillouin, \it Wave Propagation and Group Velocity \rm
(Academic, New York, 1960).

[6]. A. Enders and G. Nimtz, \it Phys. Rev. \rm \bf B 48, \rm 632
(1993).

[7]. A. Enders and G. Nimtz, \it J. Phys. I France \rm \bf 2 \rm
1693 (1992).

\end{document}